# The Bowley Ratio

## Abstract

The paper gives a simple algebraic description, and background justification, for the Bowley Ratio, the relative returns to labour and capital, in a simple economy.

## Background

*"I mean the stability of the proportion of national dividend accruing to labour, irrespective apparently of the level of output as a whole and of the phase of the trade cycle. This is one of the most surprising, yet best-established, facts in the whole range of economic statistics………Indeed… the result remains a bit of a miracle."* [Keynes 1939]

*"…no hypothesis as regards the forces determining distributive shares could be intellectually satisfying unless it succeeds in accounting for the relative stability of these shares in the advanced capitalist economies over the last 100 years or so, despite the phenomenal changes in the techniques of production, in the accumulation of capital relative to labour and in real income per head."* [Kaldor 1956]

*"FUTURE ISSUES - Theory*

*1. Is there a deep explanation for the coefficient of 1/3 capital share in the aggregate capital stock? This constancy is one of the most remarkable regularities in economics. A fully satisfactory explanation should not only generate the constant capital share, but some reason why the exponent should be 1/3 (see Jones 2005 for an interesting paper that generates a Cobb-Douglas production function, but does not predict the 1/3 exponent). With such an answer, we might understand more deeply what causes technological progress and the foundations of economic growth."* [Gabaix 2009]

The ratio of the returns to labour to total returns has been a long-standing mystery of economics. While carrying out modeling work of a simple economy, the author found that the formula for the Bowley ratio 'emerged' naturally out of the model. The nature of the formula made the underlying explanation for the Bowley ratio, and the constancy of this value apparent. This brief paper gives the derivation and the background justification.

The modelling work that gave rise to the definition of the Bowley ratio is wide ranging and gives explanations for power tails in wealth and income distributions and also company size distributions. This work is written up in the paper 'Why Money Trickles Up' which can be downloaded at econodynamics.org.

For most mature economies the ratio of returns to labour to total returns is a near constant that varies between about two-thirds and three-quarters. Young gives a good discussion of the national income shares in the US, while Gollin gives a very thorough survey of income shares in more than forty countries [Young 2010, Gollin 2002].

The constancy of the Bowley ratio is unexpected; in the long run it seems logical that mechanisation and the increasing use of capital would result in the Bowley ratio slowly moving towards zero, with more returns to capital and less to labour.

In fact if you analyse the data on a sectoral basis, this is exactly what is happening. Young shows clearly that for US agriculture and manufacturing, returns to labour have declined significantly while returns to capital have increased. In the US returns to labour in agriculture have dropped from nearly 0.8 of total income in 1958 to less than 0.6 by 1996. In manufacturing, the change has been from 0.75 to two-thirds.

This has happened because labour has been slowly displaced by machines in these industries. The fascinating thing is that despite the changes in the Bowley ratios for these two (very large) sectors, the national value of the Bowley ratio for the US as a whole has stayed near constant between 0.69 and 0.66 using the same measures.

The reason for this is that the labour intensive service sector has grown dramatically in size through the same period, and this has kept the national balance of returns to labour and capital very nearly constant.

**Derivation**

Although I discovered formula (15) below from my modelling work, it is trivial to derive the Bowley ratio algebraically. Assume an isolated economy at equilibrium, with zero growth, without a state sector, and no debt; so all values of flows and stocks are constant.

At this equilibrium point, the total capital is constant, total income must equal total consumption, so all the definitions below hold.

Here e is the earnings paid to labour, $\pi$ is the profit and can refer to any income from paper assets such as dividends, rent, coupons on bonds, interest, etc, and W is the total capital or wealth represented by the paper assets.

$$\text{Consumption} = \text{Income}$$

$$C = Y \quad (1)$$
$$C = e + \pi \quad (2) \quad \text{and:}$$

$$\text{Consumption rate} \quad \Omega = \frac{C}{W} \quad (3)$$

$$\text{Income rate} \quad \Gamma = \frac{Y}{W} \quad (4)$$

$$\text{Profit rate} \quad r = \frac{\pi}{W} \quad (5)$$

$$\text{Bowley ratio} \quad \beta = \frac{e}{Y} \quad (6)$$

$$\text{Profit ratio} \quad \rho = \frac{\pi}{Y} \quad (7)$$

$$\beta + \rho = 1 \quad (8)$$

$$\text{Profit ratio} \quad \rho = \frac{r}{\Gamma} \quad (9)$$

First multiply equation (1) by equation (5), then we get:

$$\frac{\pi}{W}C = rY \quad (10)$$

Substituting from (3) into the left hand side gives:

$$\pi\Omega = rY \quad (11)$$

Rearranging gives:

$$\frac{\pi}{Y} = \frac{r}{\Omega} \quad (12)$$

substituting from (7) gives the profit ratio:

$$\rho = \frac{r}{\Omega} \quad (13)$$

Subtracting both sides from unity gives:

$$1 - \rho = 1 - \frac{r}{\Omega} \quad (14)$$

or, substituting from (8):

$$\beta = \text{Bowley ratio} = 1 - \frac{r}{\Omega} \quad (15)$$

This is the formula for the Bowley ratio that 'emerged' from my modelling.

At this point, more observant readers may have noticed the similarity of equations (9) and (13) givng:

$$\rho = \frac{r}{\Omega} = \frac{r}{\Gamma} \quad (16)$$

which clearly means:

$$\Omega = \Gamma \quad (17)$$

from the definitions of $\Omega$ and $\Gamma$ it then follows that:

$$\frac{C}{W} = \frac{Y}{W} \qquad (18) \qquad \text{so:}$$

$$C = Y \qquad (19)$$

which of course was the original definition of (1).

**Discussion**

For most economists the above will appear to be a tautology, and a trivial and unimportant accounting identity. But it isn't. It is all a question of directionality. Of cause and effect.

For most people it is 'obvious' that consumption follows income, ie that people earn then spend, or that:

$$C = Y$$

Actually it is the other way round:

$$Y = C \qquad \text{or more accurately:}$$

$$\Gamma = \Omega$$

It is the consumption rate $\Omega$ that defines $\Gamma$; the ratio of total income to capital.

Trivially this was the case in my models, where r and $\Omega$ were fixed as exogenous, and $\Gamma$ was allowed to float. But of course this is not sufficient justification.

The problem with the economic literature with regard to the Bowley ratio is that economists have first effectively defined the profit ratio and Bowley ratio as:

$$\rho = \frac{r}{\Gamma}$$

$$\beta = 1 - \frac{r}{\Gamma}$$

They have then spent the last hundred years or so trying to explain the two ratios above by attempting to look at the microeconomic structure of industry that could affect r and $\Gamma$. This has almost entirely revolved around the analysis of 'production functions', the supposed microeconomic relations between capital and labour.

There are however major problems with this approach.
Firstly, real analysis of companies suggests that any meaningful production function needs to be based on high fixed costs and increasing returns, and is far away from the Cobb-Douglas or other standard production functions used in neoclassical economics [Keen 2004, Lee 1999].

Secondly, as the data from Young [Young 2010] shows the relative shares accruing to labour and capital can change quite significantly within individual sectors such as agriculture and manufacturing. This shows that production functions are not giving the required output on a sector-by-sector basis. (Casual inspection of company accounts also shows that returns to labour and capital can vary dramatically from company to company.)

The third and most important reason is the problems following the logical steps.

Firstly, traditional economics states that production functions define the relationship between r, the rate of return to capital, and Γ, the rate of total income to capital.

Secondly, traditional economics states that total income is equal to total consumption, so, logically, $\Omega = \Gamma$.

Putting the two statements above together gives the logical conclusion that production functions, the microeconomic structure of the commercial sector, define the saving rate $\Omega$ (leaving aside r for the moment).

This is very difficult to swallow.

Squirrels save. As do beavers. And also some woodpeckers and magpies. Almost all agricultural societies store grains and other foods to tide them from one harvest to the next. And whether you live in the tropics with alternating wet and dry seasons, or a temperate climate with warm and cold seasons, saving is a biological necessity genetically selected in human beings for its beneficial outcomes. Saving is a deeply ingrained human behaviour that borders on the compulsive.

Leaving biology aside, traditional economics has well-established logical theories for saving. Lifetime cycles make it logical for young, and especially middle-aged people to save to ensure support in their old age.

Whether you look at biology or economics, savings rates are largely exogenous to the economic system.

It stretches credulity to breaking point, to believe that saving and consumption behaviour is ultimately defined by the microeconomic production functions of commercial companies.

The causality works the other way, the systems of capitalism are set up in such a manner that the consumption rate $\Omega$ defines Γ, the rate of total income to capital.

When viewed in this way the data of Young makes sense.

In the period Young analysed, consumption rates stayed approximately constant, as did rates of return.

During the same period, both agriculture and manufacturing increased their returns to capital and reduced returns to labour.

Given fixed $\Omega$, to keep things balanced, the economy as a whole was obliged to create new, labour-intensive, industries to ensure that returns to labour were maintained as a whole.

All those cappuccino bars and hairdressers were created by the economy; by entropy, to ensure that the Bowley ratio remained equal to $1-(r/\Omega)$.

Given r and $\Omega$ as positive ratios, and that $\Omega$ is normally much larger than r, then the Bowley ratio will normally be a fraction, closer to one than zero.

It is straightforward to check equation (15) against reality. A suitable long-term profit rate could be anywhere between long-term interest rates and long-term real stock-market returns. Long-term real interest rates are generally in the region of 2% to 5% [Homer & Sylla 1996, Measuring Worth]. Long-term stock-market returns appear to be in the region of 7% to 8% [Campbell 2003, Ward 2008].

Consumption is typically about 60% of gdp [Miles & Scott 2002, section 2.2, fig 2.3]. While non-residential capital stock is typically 2.5 to 3 times gdp [Miles & Scott 2002, sections 5.1 & 14.1]. Taken together this would give Ω, the consumption rate as a proportion of capital a range of about 0.2 to 0.25.

Substituting into equation (15) this then gives a possible range of values for the Bowley ratio of between 0.60 and 0.92.

Clearly this range is a little on the high side when compared with the 'stylised facts' of observed Bowley ratios in the real world varying between the values of 0.5-0.75.

We are however in the right ballpark.

I find it difficult to believe that I am the first person to propose that the Bowley ratio should be defined by:

$$\beta = 1 - \frac{r}{\Omega} \qquad \text{rather than:}$$

$$\beta = 1 - \frac{r}{\Gamma}$$

However, I have not been able to find any other proposal of this relationship, and the recent writings of Gabaix, Young and others suggest that this is the case. If I am the first to do so I am happy to take the credit. If not I would be happy to update this paper appropriately.